\documentclass{iaus}
\usepackage{graphicx}

\title[Determination of the barycentric velocity from observations]
{Determination of the barycentric velocity of an astrometric satellite using its own observational data}

\author[A. G. Butkevich \& S. A. Klioner]
{{Alexey G. Butkevich}\index{{\bf Butkevich}, A.G.}$^1$  \thanks{On leave from Pulkovo Observatory, 196140 Saint-Petersburg, Russia.}
{Sergei A. Klioner}\index{{\bf Klioner}, S.A.}$^2$}

\affiliation{Lohrmann Observatory, Dresden Technical University,
\break 01062 Dresden, Germany \break $^1$email:
alexey.butkevich@tu-dresden.de \break $^2$email:
sergei.klioner@tu-dresden.de}

\date{20071116 and in revised form 20080121}
\volume{248}  
\pagerange{???--???}
\jname{A Giant Step: from Milli- to Micro-arcsecond Astrometry} \editors{W.J. Jin, I. Platais \& M.A.C. Perryman, eds.}
\begin{document}

\maketitle

\begin{abstract}
The problem of determination of the orbital velocity of an
astrometric satellite from its own observational data is studied. It
is well known that data processing of microarcsecond-level
astrometric observations imposes very stringent requirements on the
accuracy of the orbital velocity of the satellite (a velocity
correction of 1.45 mm/s implies an aberrational correction of 1
$\mu$as). Because of a number of degeneracies the orbital velocity
cannot be fully restored from observations provided by the satellite. Seven constraints that must be applied on
a velocity parameterization are discussed and formulated
mathematically. It is shown what part of velocity can be
recovered from astrometric data by a combined fit of both velocity
parameters and astrometric parameters of the sources. Numerical
simulations show that, with the seven constraints applied, the
velocity and astrometric parameters can be reliably estimated from
observational data. It is also argued that the idea to improve the
velocity of an astrometric satellite from its own observational data
is only useful if the a priori information on the orbital velocity
justifies the applicability of the velocity constraints. The
proposed model takes into account only translational motion of the
satellite and ignores any satellite-specific parameters. Therefore,
the results of this study are equally applicable to both scanning
missions similar to Gaia, and pointing ones like SIM, provided that
enough sources were observed sufficiently uniformly.


\end{abstract}

\firstsection 
\section{Introduction}

The astrometric accuracy of several microarcseconds announced for three space
missions (Gaia, SIM and Jasmine) implies that the velocity of the satellite should be known
within several mm/s. Such a high precision is a challenge to standard orbit
determination techniques.

Another way to proceed was discussed in the Gaia community since
2001 and formulated in a written form by \cite{klioner05}. The idea
is to use Gaia's own astrometric data to fit a correction to the
Gaia velocity. The velocity correction $\delta\mathbf v$ is the
difference between the real velocity of the satellite and its
ephemeris velocity available a priori. A straightforward theoretical
analysis of this idea (\cite[Butkevich 2006]{butkevich06})
demonstrated that velocity can be determined from observations with
the required precision, provided that all other parameters are
exactly known. Some criticism against this approach has been
formulated by \cite{bastian04a} who argued that the velocity
obtained from observation will strongly correlate with the source
parameters. A detailed theoretical exposition of this problem has
been done by \cite{klioner07}, who explicitly demonstrated that the
velocity correction to be fitted from the data must satisfy some
constraints.

\section{What part of observer's velocity can be restored from astrometric data?}\label{sec:whatpart}

Let us consider the case when the velocity correction has a constant component:
\begin{equation}\label{v0}
\delta\mathbf v=\mathbf v_0=\mathrm{const}\,,
\end{equation}
then this additional velocity leads to the following aberrational correction
\begin{equation}\label{dsv}
\delta\mathbf u\approx-\mathbf u\times\left[\mathbf u\times\mathbf v_0/c\right]\,.
\end{equation}
This correction depends on the stellar position $\mathbf u$, but
for each star it is constant. Such a signal in $\delta\mathbf v$
is  equivalent to a constant change of positions, which cannot be
detected from observations since it cannot be distinguished from
their {\it real} change.

Similarly, for a velocity correction changing linearly with time,
\begin{equation}\label{a0}
\delta\mathbf v=\mathbf a_0t\quad\mathbf a_0=\mathrm{const}\,,
\end{equation}
we have
\begin{equation}\label{dsa}
\delta\mathbf u\left(t\right)\approx\boldsymbol{\mu}_0t\,,
\end{equation}
where
\begin{equation}\label{mu}
\boldsymbol{\mu}_0=-\mathbf u\times\left[\mathbf u\times\mathbf a_0/c\right]\,.
\end{equation}
This correction is equivalent to an
additional constant proper motion for each star. Such a correction again
cannot be detected from observations since it cannot be distinguished
from a {\it real} change of the proper motion parameters for each
source.

If $\delta\mathbf v$ is exactly proportional to the barycentric
position of the satellite $\mathbf r\left(t\right)$:
\begin{equation}\label{alpha0}
\delta\mathbf v\left(t\right)=\alpha_0\mathbf r\left(t\right)\quad\alpha_0=\mathrm{const}\,,
\end{equation}
the corresponding first-order aberrational correction reads
\begin{equation}\label{dsalpha}
\delta\mathbf u\approx-\mathbf u\times\left[\mathbf u\times\alpha_0\mathbf r/c\right]\,.
\end{equation}
On the other hand, the aberrational effect caused by a global offset of parallaxes $\delta\pi$ is
\begin{equation}\label{dspi}
\delta\mathbf u=\mathbf u\times\left[\mathbf u\times\delta\pi\mathbf r/\mathrm{AU}\right]\,.
\end{equation}
These effects are indistinguishable provided that
\begin{equation}\label{pi}
\delta\pi=-\alpha_0\mathrm{AU}/c\,.
\end{equation}

Thus the problem has seven free parameters ($\alpha_0$ and six components of
$\mathbf v_0$ and $\mathbf a_0$), which correlate with some astrometric
information, i. e. it has seven degrees of freedom. This rank deficiency makes
the direct determination of velocity impossible. It can be, however,
demonstrated that the degeneracy can be eliminated if the following constraints
would be imposed onto the solution:
\begin{eqnarray}\label{c}
\int\limits_0^T\delta\mathbf v\left(t\right)\mathrm dt=0 &
\quad\mbox{to remove}\quad & \delta\mathbf v=\mathbf v_0\,, \\
\int\limits_0^T \left(t-T/2\right)\delta\mathbf v\left(t\right)\mathrm dt=0 & \mbox{to remove} &
\delta\mathbf v=\mathbf a_0t\,, \\
\int\limits_0^T\frac{\delta\mathbf v\left(t\right)\mathbf
r\left(t\right)}{\left|\mathbf r\left(t\right)\right|^2}\,\mathrm
dt=0 & \mbox{to remove} & \delta\mathbf v=\alpha_0\mathbf r \,.
\end{eqnarray}

These constraints guarantee that the solution does not contain the relevant
signals in the sense of least-squares.

Although it can hardly be proved analytically that the problem has no other
degrees of freedom, it can be checked numerically. We calculated singular value decomposition (SVD) of a normal
matrix and found only  seven small singular values shown in Table~\ref{table:1}.
This fact states that no other degrees of freedom exist.

\begin{table}\def~{\hphantom{0}}
  \begin{center}
\caption{\label{table:1} Singular values of the normal matrix.}
\begin{tabular}{ccccccc}
\hline 
$n$        & 1 & \vdots & 7 & 8 & \vdots & max \\
$\sigma_n$ &  $9.6\cdot 10^{-8}$ & \vdots & $9.0\cdot 10^{-6}$ & 2.1 & \vdots & 1045 \\
\hline 
\end{tabular}
 \end{center}
\end{table}

\section{Legitimacy of the constrained velocity}

The constraints may be only applied when a priori accuracy of the
ephemeris guarantees that no signal of given kinds can exist in
real velocity, or, strictly speaking the signal is so small that any effect
due to it can be completely neglected.

The \emph{real} $\delta\mathbf v_\mathrm{real}$ velocity correction
can be represented as a sum of the two different components
\begin{equation}\label{vt}
\delta\mathbf v_\mathrm{real}=\delta\mathbf v+\mathbf
R\left(t\right)\,,
\end{equation}
where $\delta\mathbf v$ is the component that can be fitted and
$\mathbf R\left(t\right)$ is the component violating the seven
constraints. The fitted component $\delta\mathbf v$ is useful if and
only if the uncertainty of the ephemeris velocity is such that it
guarantees that
\begin{equation}
\left|\mathbf R\left(t\right)\right|<\epsilon
\end{equation}
at any instant of time. Here $\epsilon$ is a required velocity
accuracy (for Gaia, for example, $\epsilon=1$~mm/s).

Fortunately, this can be demonstrated for Gaia (\cite[Klioner \& Butkevich, 2007]{klioner07})
but cannot be guaranteed for other missions.

\section{Results of the numerical simulations}

To study the problem of the velocity determination numerically, we
have implemented a simple simulator of Gaia observations - Dresden
Gaia Simulator (DGS) - that includes many of the basic features
of the real mission. Table~\ref{table:results} shows the results
of a simulation run with 2048 stars covering 5 years of
observations (\cite{butkevich07a}). The accuracy of an individual
observation was chosen to be 30 $\mu$as. The small errors found
suggest that the constrained solution allows one to achieve a precise
and reliable determination of velocity and source parameters.

\begin{table}\def~{\hphantom{0}}
  \begin{center}
\caption{\label{table:results} Error in parameters.}
\begin{tabular}{ccccccccc}
\hline 
Parameter & $\alpha$ & $\delta$ & $\mu_\alpha$ & $\mu_\delta$ & $\pi$ & $\delta v_x$ & $\delta v_y$ & $\delta v_z$ \\
\hline 
Error & 1.6 $\mu$as & 1.5 $\mu$as & 1.3 $\mu$as/yr & 1.2 $\mu$as/yr & 1.5 $\mu$as & 1.8 mm/s & 1.5 mm/s & 1.6 mm/s \\
\hline 
\end{tabular}
 \end{center}
\end{table}

\section{Assessment of the accuracy of velocity determination}

\begin{figure}
\centering
\includegraphics[height=108pt]{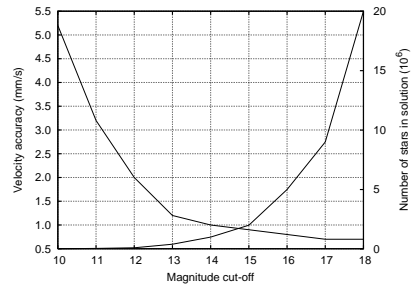}
\caption{Dependence of the velocity accuracy (the descending
curve) and the number of stars in the astrometric solution (the
ascending curve) on the limiting magnitude.} \label{figv}
\end{figure}

Our simulation has one serious drawback -- it can only
handle limited datasets and never approaches the data volume close
to the expected Gaia parameters. A straightforward estimation can
be made using simple statistical considerations
(\cite{butkevich07b}). One of the critical parameters in the
velocity determination is the number of stars used in the solution,
which depends on the apparent magnitude cut-off. The estimated
velocity accuracy together with relevant star counts are shown in
Fig.~\ref{figv} for the limiting $V$ magnitudes. Besides the number
of stars, the accuracy also depends on the temporal resolution
of the fitted velocity correction. The accuracy obviously degrades when a finer
resolution is used. The time scale of velocity variations was chosen to be
6 hours, close to the rotation period of Gaia. We may conclude from
the obtained estimates that at least $10^6$ stars are needed to
obtain velocity with an accuracy of 1 mm/s.

\section{Satellite specific parameters}

The discussed model takes into account only translational motion of
the satellite and ignores any satellite-specific calibration
parameters (e. g. attitude parameters). The situation may become more complicated
when those other parameters are also considered. 
Our analysis shows that for Gaia 
scientifically important parameters can be successfully restored even in this case.

\begin{acknowledgments}
We acknowledge useful discussions with Ulrich Bastian and Lennart Lindegren.
This work was partially supported by the BMWi grant 50\,QG\,0601
awarded by the Deutsche Zentrum f\"ur Luft- und Raumfahrt e.V. (DLR).
\end{acknowledgments}


\begin{thebibliography}{}

\bibitem[{Bastian (2004a)}]{bastian04a}
   Bastian, U. 2004a,
   Improving Gaia's orbit with Gaia's astrometry?,
   GAIA-ARI-BAS-007

\bibitem[{Butkevich (2006)}]{butkevich06}
   Butkevich, A.~G. 2006,
   On the velocity determination from observational data,
   GAIA-CA-TN-LO-AGB-001-1

\bibitem[{Butkevich \& Klioner 2007a}]{butkevich07a}
   Butkevich, A.~G. \& Klioner, S.~A. 2007a,
   On the simultaneous determination of velocity correction and source parameters,
   GAIA-CA-TN-LO-AGB-002-1

\bibitem[{Butkevich \& Klioner 2007b}]{butkevich07b}
   Butkevich, A.~G. \& Klioner, S.~A. 2007b,
   Assessing the accuracy of the velocity determination,
   GAIA-CA-TN-LO-AGB-005-1

\bibitem[{Klioner (2005)}]{klioner05}
   Klioner, S. 2005,
   On the possibility to improve the velocity of Gaia from the Gaia's own astrometric data,
   available from Gaia Livelink

\bibitem[{Klioner \& Butkevich (2007)}]{klioner07}
   Klioner, S.~A. \& Butkevich, A.~G. 2007,
   What part of observer's velocity can be restored from astrometric data?,
   GAIA-CA-TN-LO-SK-001-2

\end{thebibliography}
\end{document}